# Construction of the Complete Set of Maximally Entangled Basis Vectors for N-Qubit Systems

Chi-Chuan Hwang [1]*

[1] Department of Engineering Science, National Cheng Kung University (NCKU), Tainan, 70101, Taiwan.

* Corresponding authors: Chi-Chuan Hwang (chchwang@mail.ncku.edu.tw);

**Abstract**

In this study, we first use a three-qubit system as an example to demonstrate the construction of quantum circuits for the eight maximally entangled basis vectors, subsequently extending the approach to N-qubit systems. We employ a random-number approach to generate maximally entangled basis vectors and their corresponding circuits, while also detailing the required number of single-qubit and CNOT gates. This approach not only provides a solid theoretical foundation but also establishes a practical technique for technological applications, bypassing the difficulty of storing large-scale encoding data.

**Introduction**

In the study of quantum computing, entangled states are widely utilized for both computation and communication applications. In the case of a two-qubit system, the most well-known example of a maximally entangled basis is the set of Bell states. For three-qubit states, only specific states such as |GHZ⟩ [1] and |W⟩ [2] have been commonly used. However, systematically constructing a complete set of $2^N$ maximally entangled basis vectors for N-qubit systems remains a significant challenge in the academic community. This is precisely the problem this paper aims to resolve. The breakthrough presented in this study is expected to have a significant impact on applications in various fields, including quantum sensing and metrology [3][4], quantum communication [5][6][7][8], quantum cryptography [9][10], quantum computation and simulation [11][12], fundamental tests of physics [13][14], the construction of quantum networks [15][16], and quantum radar [17].

In the following sections, we first introduce the eight maximally entangled basis vectors for the three-qubit case. We then present a method for constructing the $2^N$ maximally entangled basis vectors for general N-qubit systems, followed by a brief conclusion.

## Construction of Maximally Entangled Vector Tracing for Three-Qubit System

First, we introduce the complete set of maximally entangled basis vectors for the three-qubit case, which consists of eight states in total. For clarity, we present the quantum circuits used to generate these basis vectors together with their corresponding derivation formulas, as illustrated in Fig.1~4.

Fig.1(a) shows the first basis vector, which initial state is $|000\rangle$. A Hadamard gate is applied to the first qubit, followed by two CNOT gates, whereas the first being $CNOT_{0,1}$ and the second $CNOT_{0,2}$. The computation proceeds in four steps, resulting in the basis vector $\frac{|000\rangle+|111\rangle}{\sqrt{2}}$, which corresponds to the state $|GHZ+\rangle$[18]. As shown in Fig.1(b), the main difference from the above basis vector is the addition of a Z quantum gate after the Hadamard gate, which yields the basis vector $\frac{|000\rangle-|111\rangle}{\sqrt{2}}$.

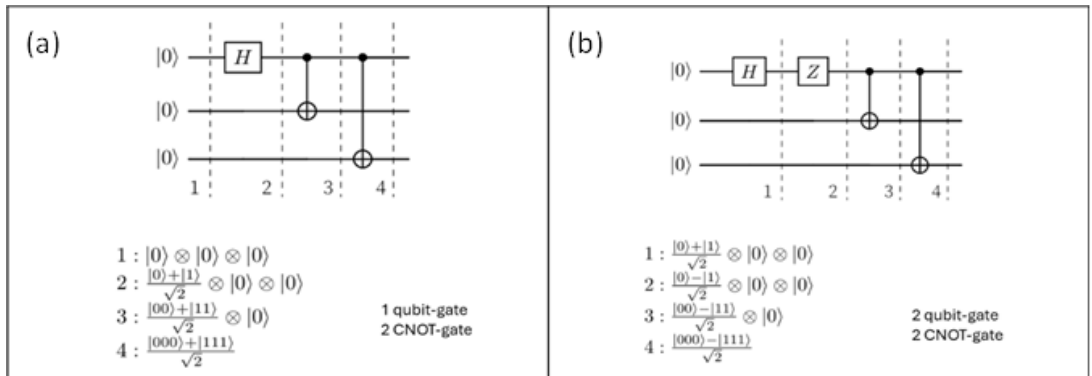

Fig.1 Quantum circuit of the basis vector $\frac{|000\rangle\pm|111\rangle}{\sqrt{2}}$ and the four-step derivation formula.

Similarly, the third basis vector is obtained by applying a Hadamard gate to the first qubit, followed by a CNOT-gate ($CNOT_{0,1}$) and then an inverted-control CNOT-gate ($CNOT^{(0)}_{0,2}$). The same calculation proceeding in four steps results in the basis vector $\frac{|001\rangle+|110\rangle}{\sqrt{2}}$, as shown in Fig.2(a). In a similar manner, adding a Z quantum gate immediately after the initial Hadamard gate will yield the fourth basis vector as $\frac{|001\rangle-|110\rangle}{\sqrt{2}}$, as illustrated in Fig.2(b).

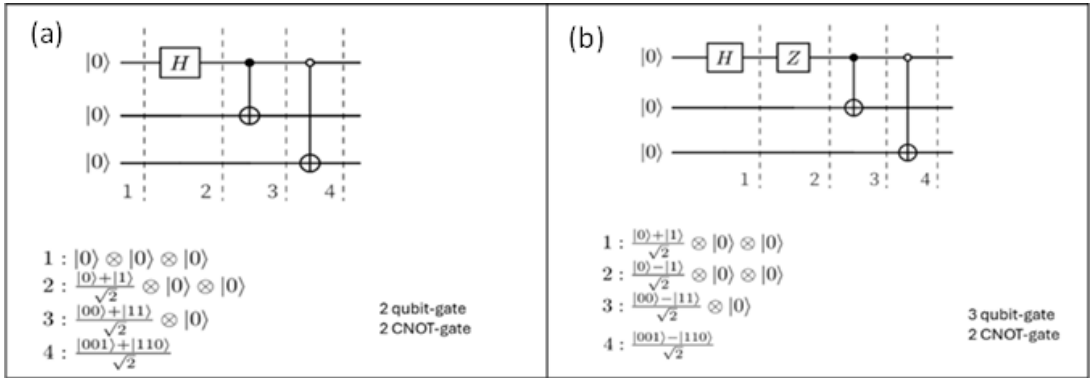

Fig.2 Quantum circuit of the basis state $\frac{|001\rangle \pm |110\rangle}{\sqrt{2}}$ and the four-step derivation formula.

Likewise, applying a a Hadamard gate to the first qubit will yield the fifth basis vector, followed by an inverted-control CNOT -gate $CNOT^{(0)}_{0,1}$, and then another CNOT- gate $CNOT_{0,2}$. The computation proceeds in four steps to yield the basis vector $\frac{|010\rangle+|101\rangle}{\sqrt{2}}$, as shown in Fig.3(a).

In a similar manner, inserting a Z gate immediately after the initial Hadamard gate will yield the sixth basis vector s $\frac{|010\rangle-|101\rangle}{\sqrt{2}}$, as illustrated in Fig.3(b).

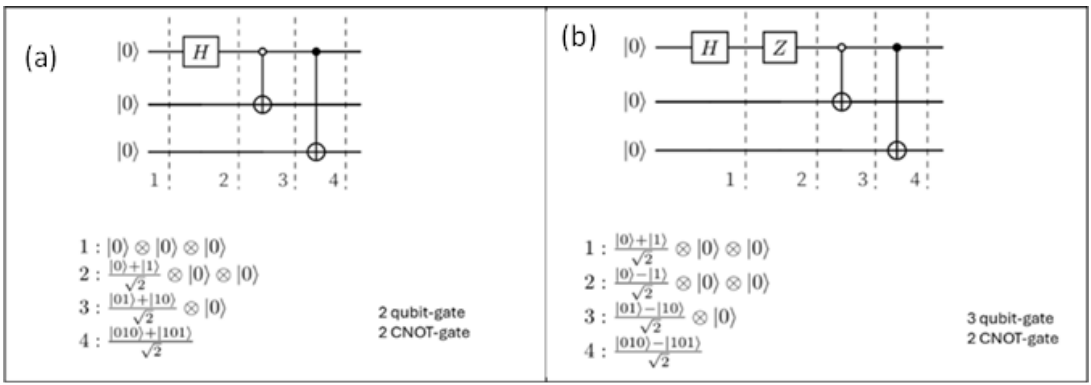

Fig.3 Quantum circuit of the basis state $\frac{|010\rangle \pm |101\rangle}{\sqrt{2}}$ and the four-step derivation formula.

Finally, applying a Hadamard gate to the first qubit will yield the seventh basis vector, followed by an inverted-control CNOT gate $CNOT^{(0)}_{0,1}$, and then another inverted-control CNOT gate $CNOT^{(0)}_{0,2}$. The calculation proceeds in four steps, resulting in the basis vector $\frac{|011\rangle+|100\rangle}{\sqrt{2}}$, as shown in Fig.4(a). In a similar manner, adding a Z quantum gate immediately after the initial Hadamard gate will yield the eighth basis vector $\frac{|011\rangle-|100\rangle}{\sqrt{2}}$, as illustrated in Fig.4(b).

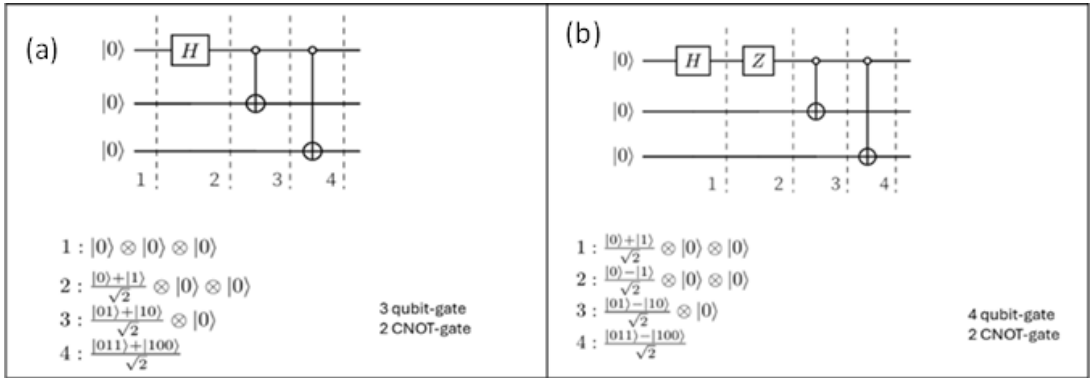

Fig. 4 Quantum circuit of the basis state $\frac{|011\rangle \pm |100\rangle}{\sqrt{2}}$ and the four-step derivation formula.

In summary, the initial vector $|0\rangle$ acts with a Hadamard gate and transforms into $\frac{|0\rangle+|1\rangle}{\sqrt{2}}$, where vector $|0\rangle$ is the first term and vector $|1\rangle$ is the second term. All subsequent maximally entangled basis vectors are arranged in this manner. We first consider the effect of vector $|1\rangle$: if it encounters a CNOT gate, it transforms the control subject $|0\rangle$ into $|1\rangle$; if it encounters an inverted control, the control subject remains $|0\rangle$. According to this rule, when the first term $|0\rangle$ encounters a CNOT gate, the control subject remains unchanged. However, if it encounters an inverted control, the control subject becomes $|1\rangle$. For the final basis vectors paired in this manner and if the latter vector is $|1\rangle$ ($|0\rangle$) on the same qubit, the former vector must be $|0\rangle$ ($|1\rangle$). If a Z quantum gate is added after the Hadamard gate, the resulting vectors are identical except for adding a negative sign in the second term. With this pairwise combination, this study examines the four possible configurations of CNOT gates and inverted controls starting from the first qubit (Index 0), as shown in Fig. 1(a)~5(a). All four (b) cases are paired with maximally entangled basis vectors containing a term with a negative sign. Thus, we can obtain the eight basis vectors with maximally entangled basis vectors in a three-qubit system.

We can infer N-Qubits from this approach. It is notwithstanding that we only need to focus on whether the first qubit is marked as ● or ○ and the position to determine the outcome for pairing the first qubit with $|1\rangle$, and subsequently infer the outcome paired with $|0\rangle$. This is the most convenient diagrammatic presentation we can adopt when constructing the maximally entangled basis vectors in an N-qubit system.

**Construction of Maximally Entangled Basis Vectors for N-Qubit Systems**

The arrangement of N qubits is shown in Fig.5, where all control qubits are placed on the first qubit (index 0), from left to right. The first control position (index 0) (● or ○) controls index 1, the second control position (index 2) controls index 2, and so forth. With index (N-1)-th position controlling the index (N-1)-th qubit, we can generate a random sequence of N-1 qubits consisting of 0s or 1s and resulting in a $2^{N-1}$ possibilities. By giving an example of one scenario, we can deduce all other possible outcomes. As shown in Fig.5, if the random sequence is, for example, 101…1, we can arrange ● at position corresponding to 1 and ○ at positions corresponding to 0. The arrangement is determined by the digits of the second term of the vector in Fig.5. The form of the maximally entangled basis vector for the N qubits is given as:

$$\frac{1}{\sqrt{2}}(|0010\ldots0\rangle + |1101\ldots1\rangle) \tag{1}$$

Different random sequences will alter the form of the above expression, and there are $2^{N-1}$ possible configurations. That is, there are $2^{N-1}$ entangled states of this form. If a Z quantum gate is added after the Hadamard gate on the first qubit, it will yield:

$$\frac{1}{\sqrt{2}}(|0010\ldots0\rangle - |1101\ldots1\rangle) \tag{2}$$

As described above, there are $2^{N-1}$ possible random sequences. Therefore, combining the two cases will yield $2^N$ basis vectors while each basis vector is mutually orthogonal. Thus, we have successfully constructed $2^N$ maximally entangled basis vectors for an N-qubit system, where the probability of measuring $|0\rangle$ or $|1\rangle$ at each qubit position is $\frac{1}{2}$.

For the first qubit, there are N-1 positions to place the random sequence, and if ● appears M times and ○ appears L times, then N−1=L+M. In the absence of a Z gate, N−1 CNOT gates and N−M identity logic gates will be required. In the presence of a Z gate, N−M+1 identity logic gates will be required.

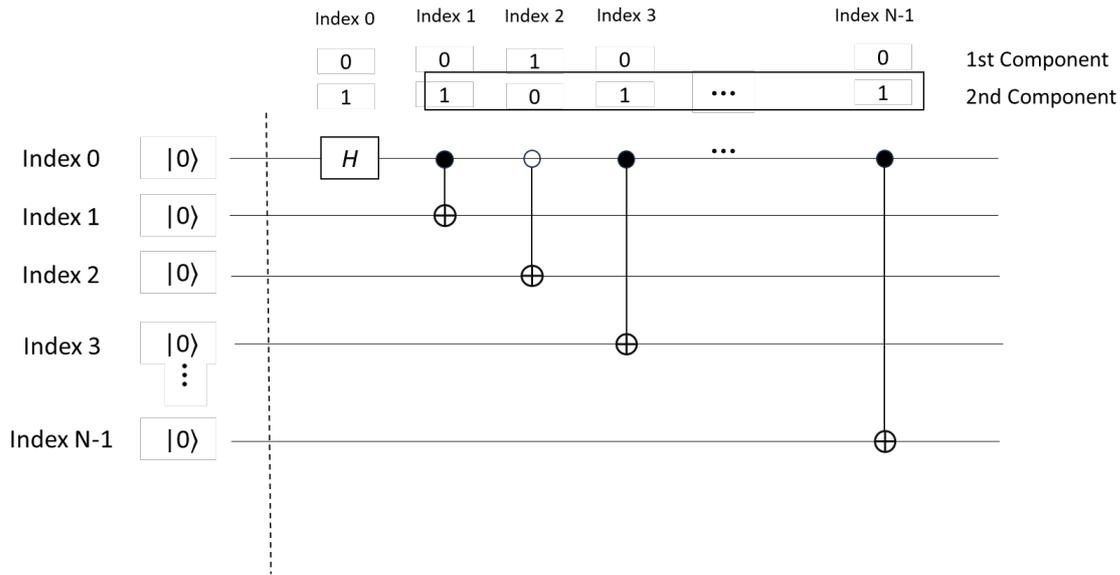

Fig.5 Quantum circuit diagram including random string 1 01…1 for generating a maximally entangled basis vector

**Conclusion**

For the three-qubit case, we have constructed the eight maximally entangled basis vectors and detailed the required number of single-qubit and CNOT gates. For an N-qubit system, the number of basis vectors, 2N, becomes enormous. We used sequences generated by a random number generator to illustrate the method; namely, when the random bit was 1, we added a CNOT logic gate (M gates), and when it was 0, we added an inverted-control logic gate (L gates, where N−1=L+M). In the absence of a Z gate, there were $2^{N-1}$ possible configurations, and such a system required N−1 CNOT-type gates and N−M single-qubit gates. If, however, an additional Z gate was applied after the Hadamard gate on the first qubit, another $2^{N-1}$ possible configurations could be generated, requiring N−1 CNOT-type gates and N−M+1 single-qubit logic gates. Once the random sequence of N−1 bits is generated, the approach proposed in this work can quickly generate the corresponding quantum circuits, since practical applications do not require encoding all basis vectors. This work not only provides a solid theoretical foundation but also offers a convenient method for practical applications.

## References

[1] N. D. Mermin, "Extreme quantum entanglement in a superposition of macroscopically distinct states", Physical Review Letters 65, 1838–1840 (1990).

[2] W. Dür, G. Vidal, J. I. Cirac, *Three qubits can be entangled in two inequivalent ways,"* Phys. Rev. A 62, 062314 (2000).

[3] Y.-C. Liu, Y.-B. Cheng, X.-B. Pan, Z.-Z. Sun, D. Pan, and G.-L. Long, Quantum integrated sensing and communication via entanglement, Phys. Rev. Appl. 22, 034051 (2024).

[4] T. Sichanugrist, H. Fukuda, T. Moroi, K. Nakayama, S. Chigusa, N. Mizuochi, M. Hazumi, and Y. Matsuzaki, "Entanglement-enhanced ac magnetometry in the presence of Markovian noise," Phys. Rev. A 111, 042605 (2025).

[5] D.-H. Kim, S. Hong, Y.-S. Kim, K. Oh, S.-Y. Lee, C. Lee, and H.-T. Lim, "Distributed Quantum Sensing with Multimode N00N States," Phys. Rev. Lett. 135, 050802 (2025).

[6] H. Shen and J. Zhang, "Entanglement-enhanced quantum metrology with neutral atom arrays," Natl. Sci. Rev. 12, nwaf149 (2025).

[7] Dave Touchette, Mark Braverman, Ankit Garg, Young Kun Ko, Jieming Mao, Andrey C. Fonseca,Maximum Separation of Quantum Communication Complexity Classes,arXiv:2505.16457(2025).

[8] Maria Stawska, Jan Wójcik, Andrzej Grudka, Antoni Wójcik,Sending absolutely maximally entangled states through noisy quantum channels,arXiv:2505.06755(2025).

[9] Yu-Chen Liu, Yuan-Bin Cheng, Xing-Bo Pan, Ze-Zhou Sun, Dong Pan, and Gui-Lu Long. Quantum integrated sensing and communication via entanglement PHYSICAL REVIEW APPLIED Received 15 April 2024; revised 28 July 2024; accepted 30 August 2024; published 23 September 2024

[10] Simone Colombo, Edwin Pedrozo-Peñafiel, Albert F. Adiyatullin, Zeyang Li, Enrique Mendez, Chi Shu, Vladan Vuletić"Time-Reversal-Based Quantum Metrology with Many-Body Entangled States Nature Physics" 14 July 2022

[11] J. I. Colless, S. J. de Graaf, M. J. Peterer, *et al.*, "Generation of Long-Range Entanglement Enhanced by Error Detection", *PRX Quantum* 6, 020331 (2025).

[12] J. Zhang, Y. Song, X. Zhang, *et al.*, "Creating and controlling global GHZ entanglement on superconducting processors", *Nature Communications* **15**, 8427 (2024).

[13] Thanaporn Sichanugrist, Hajime Fukuda, Takeo Moroi, Kazunori Nakayama, So Chigusa, Norikazu Mizuochi, Masashi Hazumi, and Yuichiro Matsuzaki, Entanglement-enhanced ac magnetometry in the presence of Markovian noise, Physical Review A, 7 April 2025.

[14] Jiahao Huang, Min Zhuang, and Chaohong Lee, Entanglement-enhanced quantum metrology: From standard quantum limit to Heisenberg limit, Applied Physics Reviews, 2 July 2024.

[15] A. Smith, B. Johnson, C. Lee, et al., "Quantum network sensing with efficient multi-partite entanglement distribution via lossy channels", arXiv preprint arXiv:2505.10148 (2025).

[16] X. Wang, L. Chen, Y. Zhao, et al., "Distribution and Purification of Entanglement States in Quantum Networks", arXiv preprint arXiv:2503.14712 (2025).

[17] M. Mathews, "A Study on Quantum Radar Technology Developments and Design Consideration for its integration," arXiv:2205.14000 (2022).

[18] Y. Yeo and W. K. Chua, *Teleportation and dense coding with genuine multipartite entanglement,"*Phys. Rev. Lett. 96, 060502 (2006).